\documentclass[12pt]{iopart}
\usepackage{graphicx}
\usepackage{iopams}  
\newcommand{\be}{\begin{eqnarray}}
\newcommand{\ee}{\end{eqnarray}}

\newcommand{\hel}{\cal{ H}}

\begin{document}

%\title[Author guidelines for IOP Publishing journals in  \LaTeXe]{Resonant chirality  in light scattered from magnetodielectric  particles}
\title{Non-zero helicity extinction in light scattered from achiral  (or chiral)  small particles located  at points of null incident helicity density}.

\author{Manuel Nieto-Vesperinas}

\address{Instituto de Ciencia de Materiales de Madrid, Consejo Superior de
Investigaciones Cient\'{i}ficas\\
 Campus de Cantoblanco, Madrid 28049, Spain.\\ www.icmm.csic.es/mnv}
\ead{mnieto@icmm.csic.es}
\vspace{10pt}
\begin{indented}
\item[]December 2016
\end{indented}

\begin{abstract}
Based on a recent unified formulation on dichorism and extinction of helicity on scattering by a small particle,  dipolar in the wide sense, magnetodielectric or not, chiral or achiral, we show that such extinction is enhanced not only at resonances of the polarizabilities, but also due  to  interference between left and right circularly polarized components of the incident wave, which contributes  with appropriate parameters of the illuminating field,   even if the particle is  achiral and is placed at points of the incident field at which  the local incident helicity density is  zero.  

This phenomenon goes beyond standard circular dichroism (CD),   and we  analyze  it  in detail  on account of the values of  the several quantities, both of the incident light and the particle, involved in the process. In addition, this interference produces a term in the helicity extinction that remarkably yields information on the real parts of the electric and/or magnetic polarizabilities, which are not provided by  CD, of which that helicity extinction phenomenon may be considered a  generalization.
\end{abstract}

%\maketitle

%\section{Introduction: file preparation and submission}
%\subsubsection{Figures.} Figures should ideally be included in an article as encapsulated PostScript files

\section{Introduction}
In recent times, the concept of circular dichroism (CD)  \cite{schellman,craig,barron1} has been extended to the  extinction by scattering (or diffraction), transmission,  and/or absorption by  nanostructures that may or may not be  chiral  \cite{menzel, govorov, zambra,kivshar, PTRSnieto}, and procedures to enhance  its weak signal from absorbing molecules has been proposed  by either enhancing the helicity of the illuminating field  \cite{tang1}, interposing a resonant particle, either chiral or achiral, between the molecule near field and the detecting tip \cite{klimov,aitzolDionne,dionne};  or reinforcing CD from nanostructures by creating near field hot spots between sets of plasmonic nanoparticles according to the choice of incident polarization \cite{wang}; by making  thermal-controlled chirality in  a hybrid THz metamaterial with $VO_2$  inclusions \cite{lv},  or by  fostering the interplay between electric and magnetic dipoles of the excited molecule \cite{LiHu}.  Also a helicity optical theorem  (HOT) has recently been established  \cite{nieto1} showing that dichroism phenomena are particular effects resulting from a fundamental law of electrodynamics: the  conservation of electromagnetic helicity, \cite{lipkin,tang1,cameron1,bliokh1,norris1,norris2,corba},   also  lending sufficient conditions to produce chiral fields by scattering \cite{PTRSnieto}, and providing  answers  to   long-standing questions on the interplay  between the chirality of fields and that of matter \cite{andrews1,andrews2}. 

   The {\it helicity}  of quasi-monochromatic, i.e.  time-harmonic, light waves, which are those addressed in this paper, is equivalent to their {\it chirality} \cite{Barnett3}. The latter being a term employed  in \cite{tang1} and that, having subsequently became of widespread use in the literature, we shall also consider here.  As  stated in \cite{nieto1}, for these  time-harmonic fields both magnitudes just differ by a factor: the square of the wavenumber. However, for general time-dependent fields both quantities have  a different physical nature and hence are not equivalent. This distinction being important  in matter. Also, as noticed in \cite{cameron1}, the helicity has dimensions of angular momentum whereas the chirality does not.

In this paper we exploit the equations for the extinction of helicity and energy that we established in a previous work,  where  a unified formulation of helicity extinction and dichroism  beyond the CD concept, was put forward. Hence we now show that  CD may be generalized to 3D polarized  fields, for which we introduce a {\it helicity extinction factor g}, a particular case of which is the standard CD {\it dissymmetry factor}. In addition, we  further analyze  the extinction of helicity  on scattering of 3D polarized  fields possessing a longitudinal component, and whose projection  in the plane  transversal to the propagation direction has elliptic polarization, (namely is the sum  of a left circular (LCP) and a  right circular (RCP) wave). This helicity extinction may be generated not only,  as in  CD, by the cross electric-magnetic polarizability that characterizes the particle  chirality, or  by the incident helicity density, but also by an interference factor that mixes  the LCP and RCP components; a phenomenon  in which the above mentioned cross-polarizability plays no role, and whose existence was already shown in \cite{PTRSnieto}. We shall study it in detail here.

In this way, we discuss how $g$ assesses the  helicity extinction in comparison with that of energy. We analyze this  under different values of  the polarizabilities of a particle that we initially assume of rather general characteristics; namely, bi-isotropic, magnetodielectric and  chiral, (we shall later relax this latter property) in the resonant regions of its polarizabilities. In addition we analyze this extinction for different local values of  the incident helicity density; also assessing  the contibution of the aforementioned  interference to this  helicity extinction, in comparison with that of the particle chirality and the incident helicity density, as well as of the  polarizability resonances, that we have so  chosen in this study in order to enhance these effects.

Among the illuminating fields  whose electric and magnetic vectors fulfill the conditions leading to this interference effect, addressed in Sections 2 and 3, we shall use a Bessel beam, which has been  well studied and is know to be accessible and employed in many experiments. We shall give some of its details in Section 4, specially in connection with its functional contribution to the densities of incident energy and helicity, as well as to the aforementioned interference phenomenon between LCP and RCP components. Then in Section 5  we shall illustrate how the discussed quantities depend on the transversal position of the particle within the incident beam.

\section{3D polarized fields with LCP and RCP transversal polarization}
We address the spatial parts ${\bf E}$ and ${\bf B}$ of the electric and magnetic vectors of quasimonochromatic  fields  in their complex representation. Their scattering in a medium of  refractive index $n=\sqrt{\epsilon\mu}$  by  a  particle, that we generally consider magnetodielectric, {\em chiral} and bi-isotropic, dipolar in the wide sense \cite{sersic,nieto2}, is thus characterized by its polarizabilities, that for e.g.   a sphere are:  $\alpha_{e}=i\frac{3}{2k^{3}}a_{1}$,  $\alpha_{m}=i\frac{3}{2k^{3}}b_{1}$,  $\alpha_{em}=i\frac{3}{2k^{3}}c_{1}$,  $\alpha_{me}=i\frac{3}{2k^{3}}d_{1}=-\alpha_{em}$.    $k=n\omega/c=2\pi n/\lambda$. Where $a_{1}$, $b_{1}$ and  $c_{1}=-d_{1}$ stand for  the electric, magnetic, and magnetoelectric first Mie coefficients, respectively \cite{bohren}.  

The electric and magnetic dipole moments, $ {\bf p}$ and  ${\bf m}$, induced in the particle by this incident field are:
\be
 {\bf p}=\alpha_{e} {\bf E}-\alpha_{me}{\bf  B},\,\ \,\,\,\,
{\bf m}=\alpha_{me}{\bf E}+\alpha_{m}{\bf B}. \label{consti}
\ee
Based on the angular spectrum decomposition of optical wavefields into LCP (sign: $+$; the notation of \cite{jackson}  is followed) and RCP  (sign: $-$)  plane wave components that we established in \cite{PTRSnieto}, (we must remark that we have recently found that  this representation was also reported in  \cite{bliokh1}), both the incident and the scattered fields may be  decomposed into the sum of an  LCP and an RCP 3D wavefield. Then we address incident fields ${\bf E}$ and  ${\bf B}$, (which we shall subsequently consider to be optical beams),  expressible  as  the sum of 3D polarized fields whose transversal polarization is LCP and RCP, respectively,  thus  holding:
\be
{\bf E}({\bf r})={\bf E}^{+}({\bf r})+{\bf E}^{-}({\bf r});\,\,\,\,\,
{\bf B}({\bf r})={\bf B}^{+}({\bf r})+{\bf B}^{-}({\bf r})
=-ni[{\bf E}^{+}({\bf r})-{\bf E}^{-}({\bf r})]; \,\,\,\,\,\,\,\,\,\, \label{ang masmen}
\ee
by which   we express the dipolar moments as 
\be
{\bf p}({\bf r})={\bf p}^{+}({\bf r})+{\bf  p}^{-}({\bf r}),  \,\,\,
{\bf m}({\bf r})={\bf m}^{+}({\bf r})+{\bf m}^{-}({\bf r}). \,\,\,\,\,\label{plusmin}
\ee
 with 
\be
{\bf  p}^{\pm}({\bf r})=(\alpha_{e} \pm n i \alpha_{me}) {\bf E}^{\pm}({\bf r});\, 
{\bf  m}^{\pm}({\bf r})=(\alpha_{me} \mp n i \alpha_{m}) {\bf E}^{\pm}({\bf r}) \label{inE}. 
\ee
  The 3D polarized  wavefields of  Eq. (\ref{ang masmen}) {\it are not} just plane waves or transversally polarized beams. In a $XYZ$-Cartesian framework,  ${\bf E}$ and  ${\bf B}$ [cf. Eq. (\ref{ang masmen})] have, in general,  a $z$-component, while that in the $XY$-plane  is elliptically polarized. As  shown below, we  illustrate these electromagnetic fields by the sum of two beams propagating along $OZ$:  LCP and RCP, respectively; both circular polarizations holding  in the $XY$-plane transversal to the beam  $z$-axis. In addition, both beams have a Cartesian component along $OZ$. 

The relationship (\ref{ang masmen}) between ${\bf B}$ and ${\bf E}$ is essential for the effects that we next obtain. This is the reason why we choose, among other possiblities, an illuminating Bessel beam in Section 4.

\section{The extinction of incident helicity and energy on scattering. Beyond circular dichroism}
Using a Gaussian system of units, the  {\it densities of  helicity}, ${\cal H}$,  -or $k^{-2}$ times the {\it chirality}-,    and   {\it energy}, ${\cal W}$, (understood as a time-averaged in this work),  of  this incident field are: 
\be
{\cal H}({\bf r})= (\epsilon/2k)[|{\bf E}^{+}({\bf r})|^2 -  |{\bf E}^{-}({\bf r})|^2], \label{hel}
\ee
(see e.g. \cite{PTRSnieto,bliokh1}), and 
\be
 {\cal W}({\bf r})= (\epsilon/8\pi)[|{\bf E}^{+}({\bf r})|^2 + |{\bf E}^{-}({\bf r})|^2], \label{en}
\ee
respectively. In what follows $\Re$ and $\Im$ denote real and imaginary parts, respectively.

 The  HOT that expresses the {\it conservation of helicity}  is \cite{nieto1}
\be
\frac{2\pi c}{\mu} \Re \{ -\frac{1}{\epsilon} {\bf p}  \cdot {\bf B}^{*} +\mu  {\bf m}\cdot  {\bf E}^{*}  \}=\frac{8\pi c k^3}{3 \epsilon} \Im[{\bf p} \cdot {\bf m}^{*}]+{\cal W}_{\hel}^{a}. \,\,\,\,\label{todip4}
\ee
The left side  of (\ref{todip4}) constitutes the {\it extinction of helicity} of the incident wave on scattering with the particle. This extinction is shown in the right side of (\ref{todip4}) to be divided up into  the {\it  total helicity scattered or radiated} by the object, (i.e. the first term in this right side)  and the rate of {\it  helicity dissipation} ${\cal W}_{\hel}^{a}$  (see Eqs. (8), (11) and (12) of \cite{nieto1}), or  {\it converted helicity}, (see Sections 3.2, 3.3 and 4 of \cite{norris2}), on interaction with the scattering body.

As shown by the right side of (\ref{todip4}), as the light interacts with the particle  such extinction may convey a selective dissipation of helicity ${\cal W}_{\hel}^{a}$ which adds to a resulting total helicity of the scattered field. This latter fact agrees with \cite{norris1, norris2}.

We should recall the  analogy of the HOT with the well-known  standard optical theorem (OT) for energies 
\be
\frac{\omega}{2} \Im[ {\bf p}  \cdot {\bf E}^{*}   +{\bf m}\cdot  {\bf B}^{*}] =
\frac{c k^4}{3 n} [\epsilon^{-1} |{\bf p}|^2 + \mu |{\bf m}|^{2}] +{\cal W}^{a} . \label{top}
\ee
The left side  of (\ref{top}) is  the energy extinguished  from the illuminting field, or  rate of energy excitation in the scattering object. The first term in the right side constitutes the  total energy scattered by the dipolar object,  whereas ${\cal W}^{a}$ stands for the rate of energy absorption by the object from the illuminating wave. 

On employing Eqs.(\ref{consti})-(\ref{inE}),  the extinction of incident helicity [cf. Eq.(\ref{todip4})]: $(2\pi c/\mu) \Re \{ -\frac{1}{\epsilon} {\bf p}  \cdot {\bf B}^{*} +\mu  {\bf m}\cdot  {\bf E}^{*}  \}$, which henceforth  we denote as ${\cal W}_{\hel}^{ext}$, is expressed as \cite{PTRSnieto}
\be
{\cal W}_{\hel}^{ext}({\bf r}) \equiv   \frac{2\pi c}{\mu}\{\Im\{[{\bf p}^{+}({\bf r})+in {\bf m}^{+}({\bf r})] \cdot {\bf E}^{+\,*}({\bf r})\,\,\,\,\,\,\,\,\,\,\,\,\,\, \nonumber \\
-[{\bf p}^{-}({\bf r})-in {\bf m}^{-}({\bf r})] \cdot {\bf E}^{-\,*}({\bf r}) \}  \,\,\,\,\,\,\,\,\,\,\,\,\,\,\, \,\,\,\,\,\,\,\,\,\,\,\,\,\,  \nonumber \\
+2\Re\{\alpha_{e} -n^{2}\alpha_{m}\} 
\Im\{ {\bf E}^{-}({\bf r}) \cdot {\bf E}^{+\,*}({\bf r})\}\}  \,\,\,\,\,\,\,\,\,\,\,\,\,\,\,  \nonumber \\
= \frac{2\pi c}{\mu}\{\frac{2k}{\epsilon}\Im\{\alpha_{e}+n^2 \alpha_{m}\}{\cal H}({\bf r})
+16 \pi \sqrt{\frac{\mu}{\epsilon}} \Re\{\alpha_{me}\} {\cal W}({\bf r}) \,\,\,\,\,\,\, \nonumber \\
+2\Re\{\alpha_{e} -n^{2}\alpha_{m}\}
 \Im\{ {\bf E}^{-}({\bf r}) \cdot {\bf E}^{+\,*}({\bf r})\}\};\,\,\,\,\,\,\,\,\, \label{g}
\ee
whereas  Eqs.(\ref{consti})-(\ref{inE}) yield for the extinction of  incident energy  [cf.Eq.(\ref{top})]:  $(\omega/2) \Im[ {\bf p}  \cdot {\bf E}^{*}   +{\bf m}\cdot  {\bf B}^{*}]$, which we write as ${\cal W}^{ext}$ \cite{PTRSnieto}, (a slightly different notation is used for this quantity in \cite{PTRSnieto}): 
\be
{\cal W}^{ext}({\bf r})\equiv \frac{\omega}{2} \{\Im\{[{\bf p}^{+}({\bf r})+in {\bf m}^{+}({\bf r})] \cdot {\bf E}^{+\,*}({\bf r})\,\,\,\,\,\,\,\,\,\,\,\,\,\, \nonumber \\
+[{\bf p}^{-}({\bf r})-in {\bf m}^{-}({\bf r})] \cdot {\bf E}^{-\,*}({\bf r}) \}\nonumber \\
+2\Im( \alpha_{e} -n^{2}\alpha_{m} )
\Re \{ {\bf E}^{-}({\bf r}) \cdot {\bf E}^{+\,*}({\bf r})\}\}\nonumber \\
= \frac{\omega}{2}\{ \frac{8\pi}{\epsilon} \Im\{\alpha_{e} 
+n^2 \alpha_{m}\} {\cal W}({\bf r})+4k \sqrt{\frac{\mu}{\epsilon}}\Re\{\alpha_{me}\} {\cal H}({\bf r})\,\,\,\,\,\,\, \nonumber \\
+2\Im\{\alpha_{e} -n^{2}\alpha_{m}\}
 \Re\{ {\bf E}^{-}({\bf r}) \cdot {\bf E}^{+\,*}({\bf r})\} \}. \,\,\,\,\,\,\,\,\,\,\, \label{g1a}
\ee
In (\ref{g}) and (\ref{g1a})    ${\bf r}$ is the position vector of the center of the particle immersed in the illuminating wavefield.  Eqs. (\ref{g}) and (\ref{g1a}) are fundamental  as they establish the connection of  the  extinction of  helicity ${\cal W}_{\hel}^{ext}$ and  energy ${\cal W}^{ext}$ of the incident wave  with the densities of incident helicity ${\cal H}$ and energy ${\cal W}$, and  with the chirality of the dipolar particle, characterized by  $\alpha_{me}$. They remarkably show how the incident  ${\cal H}$ and  ${\cal W}$ contribute  to  ${\cal W}_{\hel}^{ext}$ and   ${\cal W}^{ext}$ with their roles exchanged with respect to the polarizability factors in the corresponding term where they appear. 

Notice  from the right side of (\ref{todip4}) that ${\cal W}_{\hel}^{ext}$  contains both the total scattered helicity and the incident helicity dissipation (or conversion). Similarly, from  (\ref{top})  one sees that ${\cal W}^{ext}$ contains  the total scattered energy as well as the dissipation of incident energy  in the particle. In particular, if both ${\cal W}_{\hel}^{a}$ and ${\cal W}^{a}$ are zero,  ${\cal W}_{\hel}^{ext}$  and  ${\cal W}^{ext}$ represent the total scattered helicity and energy, respectively.

Since we are here interested in the rate of extinction of helicity, we observe in (\ref{g})  that ${\cal W}_{\hel}^{ext}$, apart from being due to the incident helicity density  ${\cal H}$ coupled with the dissipative part of the electric and magnetic polarizabilities, is generated by a coupling of the incident energy density ${\cal W}$  with the particle chirality through $\Re\{\alpha_{me}\}$. Moreover, of special importance is that, as  shown by the third  term $\Re\{\alpha_{e} -n^{2}\alpha_{m}\} \Im\{ {\bf E}^{-}\cdot {\bf E}^{+\,*}\}$ in  Eq. (\ref{g}), placing the small particle at a position ${\bf r}_0$ in the illuminating wave,  an   incident field with no helicity density at ${\bf r}_0$ may give rise to an extinction rate of helicity on interaction with the particle, not only  -as well-known-  due to the particle chirality through the term with  $\Re\{\alpha_{me}\} {\cal W}$, but also, and this is the new feature addressed in this work,  because of the  interference coupling factor $ {\bf E}^{-}\cdot {\bf E}^{+\,*}$. I.e. a non-zero ${\cal W}_{\hel}^{ext}$ will be generated  at ${\bf r}_0$ even if  the incident helicity ${\cal H}({\bf r}_0)=0$ and the particle is not chiral ($\alpha_{me}=0$). Moreover,  as $\Re\{\alpha_{e}\}$ and $\Re\{\alpha_{m}\}$ are usually larger than their imaginary counterparts at non-resonant $\lambda$,  this interference term acquires special importance for molecules \cite{PTRSnieto,nieto1}.

In this respect, and in contrast with the above argument for ${\cal W}_{\hel}^{ext}$  if ${\cal H}=0$,  there is no analogous reasoning for a non-zero  energy extinction  ${\cal W}^{ext}$, Eq. (\ref{g1a}),  if the incident electromagnetic energy ${\cal W}$ is null, since  this  would convey that  ${\bf E}^{-}= {\bf E}^{+}={\cal H}=0$ and thus ${\cal W}^{ext}=0$, as it should.

However,  in (\ref{g}) $\Re\{\alpha_{e}\}$ and $n^2 \Re\{\alpha_{m}\}$ appear in  a substraction,  and thus compete with each other  in their  contribution to the last term of (\ref{g}), which becomes zero  when  $\alpha_{e}=n^{2}\alpha_{m}$, namely when the particle is {\it dual} \cite{nieto1}.

Notice that similar arguments exist for  $\Im\{\alpha_{e}\}$ and $\Im\{\alpha_{m}\}$ in  connection with the third term of the energy extinction, Eq.  (\ref{g1a}). It is also worth remarking that when $\{ {\bf E}^{-}\cdot {\bf E}^{+\,*}\}=0$, Eqs.  (\ref{g}) and  (\ref{g1a})  reduce to those standard of the HOT and the OT, respectively, (see e.g. Section V of  \cite{nieto1}).

Eqs. (\ref{g}) and (\ref{g1a}) govern a {\it generalized dichroism}  phenomenon and hence account for CD as a particular case. Namely, by defining the ratio
\begin{equation}
 g= {\cal W}_{\hel}^{ext}/{\cal W}^{ext}, \label{helifactor}
\end{equation}
which we shall  name the {\it  helicity extinction factor}, it is straightforward to see that  either  when the particle is dual,  or when the interference terms of these equations vanish like  e.g.  for elliptically polarized   plane  waves, (for which  $ {\bf E}^{-}\cdot {\bf E}^{+\,*}=0 $ since then no longitudinal $z$-component exists), then  choosing as usually done:  $|{\bf E}^{+}|^2 = |{\bf E}^{-}|^2$, one has $g=\sqrt{ \epsilon/\mu}\,\lambda \,g_{CD}$, where  $g_{CD}=2({\cal W}^{ext\, +}-{\cal W}^{ext\, -})/({\cal W}^{ext\, +}+{\cal W}^{ext \, +})$ is the the well-known  {\it dissymmetry factor} of standard   CD which from  (\ref{g1a}) results in the well-known expression in terms of the particle polarizabilities \cite{schellman,craig,barron1,tang1}:  $g_{CD}=4n \Re \{\alpha_{me}\}/(\Im\{\alpha_{e}\} +n^2  \Im\{\alpha_{m}\})$.

Hence the CD phenomenon is one of the several consequences of the HOT and thus of the conservation of electromagnetic helicity. Namely,  while standard CD is observed by illuminating the particle, or structure, with a LCP plane wave only, and separately with a RCP  one; subsequently substracting the corresponding  scattered energies as: $\sqrt{ \epsilon/\mu}({\cal W}^{ext \, +}-{\cal W}^{ext \, -}$); CD may identically be observed on a unique illumination by a wave of the form (\ref{ang masmen}) with no longitudinal component along the $OZ$-propagation direction,  (e.g. a plane wave), linearly polarized in the  (transversal)  $XY$- plane, (or generally with  $|{\bf E}^{+}|^2 = |{\bf E}^{-}|^2$), therefore  whose  LCP and RCP  components   do not interfere with each other; ie. $ {\bf E}^{-}\cdot {\bf E}^{+\,*}=0 $. The extinction of helicity, normalized to the wavelength $\lambda$,  is identical to the above mentioned difference of LCP and RCP energies.

We should  remark that  the HOT also account for the   illumination of an object with  those so-called superchiral fields produced by the superposition of two counterpropagating CPL plane waves of amplitudes $E_1$ and $E_2$ of opposite helicity  as put forward in \cite{tang1}, (which on the basis of recent studies \cite{andrews1,andrews2} we prefer to  call fields enhancing the dissymmetry factor). However it is  known \cite{choi} that this method is limited to particles -or molecules- with  $\alpha_{m}\simeq 0$, because in such configuration  $g_{CD}= 4\epsilon  \Re\{\alpha_{me}\}/[ \Im\{\alpha_{e}\}(E_1-E_2)/(E_1+E_2) +n^2   \Im\{\alpha_{m}\}(E_1+E_2)/(E_1-E_2)]$. So that when  $\Im\{ \alpha_{m}\}=0$ the usually extremely small  dissymmetry factor of standard CD, (often as small as $10^{-3}$ for molecules), may be enhanced, as seen from this latter expression of $g_{CD}$,    just by choosing $E_1 \simeq E_2$, as proposed in \cite{tang1}, (or by making  $E_1 \simeq  - E_2$ when  $ \Im\{\alpha_{e}\}=0$); but it is evident that these choices of $E_1$ and $E_2$ cannot enhance $g_{CD}$ if  both $  \Im\{\alpha_{e}\} $ and $ \Im\{\alpha_{m}\}$ are non-zero.

It is remarkable that  term $2\Re\{\alpha_{e} -n^{2}\alpha_{m}\}  \Im\{ {\bf E}^{-}({\bf r}) \cdot {\bf E}^{+\,*}({\bf r})\}$ of (\ref{g}) is {\it the only one} among those expressing the extinction of helicity, Eq. (\ref{g}), that {\it provides information on the real parts of  the polarizabilities}. No other term of (\ref{g}) contain such information. In fact, there is a well-known  lack  of this information in the aforementioned  dissymmetry factor of standard CD, addressed above. Therefore, this paper shows how creating experimental conditions for this interference factor to exist, provides a source of information on $\Re\{\alpha_{e}\}$ and $\Re\{\alpha_{m}\}$ through the extinction of helicity, Eq. (\ref{g}), and  its associated  {\it extinction  factor} $g$.

\section{Illustration with a  Bessel beam}
The contribution of   $2\Re\{\alpha_{e} -n^{2}\alpha_{m}\} \Im\{ {\bf E}^{-} \cdot {\bf E}^{+\,*}\}$ in (\ref{g}) to the   helicity extinction,  while  $\Re\{ {\bf E}^{-} \cdot {\bf E}^{+\,*}\}=0$ in (\ref{g1a}), thus increasing the helicity extinction factor  $g$, is next illustrated with an  incident  beam propagating along $OZ$, elliptically polarized in the $XY$-plane, and with longitudinal component along the $z$-propagation direction  \cite{PTRSnieto}. In this case $\partial_z \simeq ik_z$;  the wavevector being written in Cartesian components as ${\bf k}=({\bf K}, k_z)$;  $K=\sqrt{k_x^2 +k_y^2}$, $k_z=\sqrt{k^2 -K^2}$). The electric vector is expressed in terms of the vector potential $\bf A^{\pm}$ \cite{allen2,babiker} as:
\be
{\bf E}^{\pm}=ik_z{\bf A}^{\pm}+ \frac{i}{k_z}\nabla (\nabla\cdot {\bf A}^{\pm}). \label{hazgen1} \\
{\bf A}^{\pm}({\bf r})=\frac{1}{ik_z}(\hat{\bf x}\pm i\hat{\bf y})u({\bf r}) e^{ik_z z}.  \label{hazgen2} \\ 
u({\bf r})=u_0^{\pm}(R,z)e^{il\phi}. \, \,\,\, \, \, \,\,R=\sqrt{x^2 +y^2}.    \label{hazgen3} \,\,\,\, \,\,\, \, \,\,\, \,\,\, \, \,\, \,\,\, \, \,\,\, \,\,\, \,\,\,\, \,\,\, \, \,\,\, \,\,\, \, \,\, \,\,\, \, \, 
\ee
So that using $\nabla(\nabla\cdot{\bf A}^{\pm})\simeq ik_z( \nabla\cdot{\bf A}^{\pm})\hat{\bf z}$ one has
\be
{\bf E}^{\pm}=e^{ik_z  z}[(\hat{\bf x}\pm i\hat{\bf y})u +\frac{i\hat{\bf z}}{k_z}(\partial_x u \pm i\partial_y u)].   \label{Egen1} \\ 
{\bf B}^{\pm}=\mp n i {\bf E}^{\pm}.\,\,\,\,\,\,\,\,\, \label{Egen2}
\ee
Which fulfills  both $\nabla \cdot  {\bf E}^{\pm}=0$ and Eq. (\ref{ang masmen}). We shall address  the Bessel function of integer order: $u_0^{\pm}(R,z)=e_0^{\pm} J_l(KR)$, ($e_0^{\pm}$ being  constant  amplitudes)  which,  from  (\ref{hazgen1}) -  (\ref{Egen2}) and after a calculation using the recurrence relation: $J_{l-1}(x)+J_{l+1}(x)=(2l/x)J_l(x)$, leads to the form (\ref{ang masmen})  for a Bessel beam,  whose components ${\bf E}^{\pm}$ are  LCP and RCP, respectively,  in the $XY$-plane transversal to its $z$-direction of propagation. Viz.:
\be
{\bf E}^{\pm}({\bf r})=e_{0}^{\pm}e^{i(k_z z+l\phi)} [J_{l} (KR)(\hat{\bf x}\pm i\hat{\bf y})\,\,\,\,\,\,\,\,\,\,\,  \nonumber \\
 \mp \frac{iK}{k_z}   \exp(\pm i\phi)J_{l\pm 1} (KR)\hat{\bf z}]. \label{bessel}
\ee
Eq. (\ref{bessel}) coincides  with those of \cite{allen2,babiker,volke} characterizing Bessel beams.   We have nevertheless undertaken here the derivation of this kind of beams  from a first basis in order to guarantee that this field fulfils the important condition (\ref{ang masmen}), [(see Eqs. (\ref{Egen1}) and (\ref{Egen2})].

From (\ref{bessel})  we obtain
\be
|{\bf E}^{+}({\bf r})|^2 \pm |{\bf E}^{-}({\bf r})|^2=2(|e_{0}^{+}|^2 \pm |e_{0}^{-}|^2)J_l^2(KR)\,\,\,\,\,\,\,\,\,\,  \nonumber \\
+ \frac{K^2}{k_z^2}[J_{l+1}^2(KR)|e_{0}^{+}|^2 \pm |e_{0}^{-}|^2
J_{l-1}^2(KR)]. \,\,\, \, \,\,\, \,\,\, \, \,\, \,\label{sumdifbess}
\ee
On the other hand, the factor  $ \Re (\Im)\{ {\bf E}^{-}\cdot {\bf E}^{+\,*}\}$  reduces to the contribution of the field $z$-component:
\be
 \Re ( \Im)    \{{\bf E}^{-}\cdot {\bf E}^{+\, *}\}= \Re  (\Im)  \{{ E}_{z}^{-}\cdot { E}_{z}^{+\,*} \} \,\,\,\,\,\, \, \,\,\, \,\,\, \, \,\, \, \, \,\,\, \,\,\, \, \,\, \,\nonumber \\
=-\frac{K^2}{k_z^2}J_{l-1}(KR)J_{l+1}(KR)
 \Re  (\Im) \{e_{0}^{-}e_{0}^{+ \, *}e^{-2i\phi} \}
\,  . \,\,\, \, \,\,\, \,\,\, \, \,\, \,\label{sumdif}
\ee
Therefore, either of these quantities, $\Re[\cdot]$ or $\Im[\cdot]$, may  be made arbitrarily small (or zero) depending on the choice of parameters  $e_{0}^{-}$ and $e_{0}^{+}$  for the beam in the factor $  \Re ( \Im) \{e_{0}^{-}e_{0}^{+ \, *}e^{-2i\phi} \}$.  Since according to \cite{volke}, (see Fig. 6 in this reference), this beam rotates a particle of diameter $1\mu m$ to $6\mu m$ placed in the inner ring of maximum intensity in about 16 s per revolution, we shall assume the signal detection time large enough for the azimuthal angle $\phi$   of the particle center position  ${\bf r}$ not to contribute to this factor,  so that we just  consider the quantity $  \Re ( \Im) \{e_{0}^{-}e_{0}^{+ \, *}\}$.  Hence  choosing for example $e_{0}^{-}/e_{0}^{+} = \pm a  \exp(i b \pi/2) $, $a$ and $b$ being  real, the value of $  \Re (\Im)\{{\bf E}^{-}\cdot {\bf E}^{+\, *}\}$ will oscillate about zero as \, $ \cos (b \pi/2) \, ( \sin (b \pi/2) )$.

Also, depending  on the choice of the position ${\bf r}$ of the particle in the beam, and thus of  the argument $KR$, one will have in Eqs. (\ref{g}) and (\ref{g1a}) the third terms, whose  $\Re (\Im)\{{\bf E}^{-}\cdot {\bf E}^{+\, *}\}$   factor is given by (\ref{sumdif}),  comparable, or not,  to the first and second terms  whose  $(|{\bf E}^{+}({\bf r})|^2 \pm |{\bf E}^{-}({\bf r})|^2)$  factor is (\ref{sumdifbess}). This is seen observing the factor $(K^2 / k_z^2)J_{l-1}(KR) J_{l+1}(KR)$ in (\ref{sumdif}) which may be made either much larger or smaller than the term of $J_l^2(KR)$ which is the dominant contribution to (\ref{sumdifbess}).  

This latter important fact  will be seen in Section 5 by choosing two different positions of the particle in the beam, i.e. two distinct values of $KR$.

\section{Example: Enhancements in the extinction of helicity on scattering  with a resonant particle, aither chiral or not}

 To better illustrate these effects we address them at resonant wavelengths,  so that there is field enhancement on interaction with the particle, which in principle  we consider generally magnetodielectric   and chiral.  We shall later relax the latter property. We have found in the recent work \cite{silvei} a  particle model with these characteristics, and thus we consider it useful for our illustration.  Its  linear dimension  is not larger than $204$ nm. (See details of this  particle, made of a composite metal (silver)-dielectric in vacuum, $n=1$, in Fig. 3 of  \cite{silvei}). Both helicity dissipation, or conversion, ${\cal W}_{\hel}^{a}$, and  energy absorption ${\cal W}^{a}$, are  susceptible of  taking place, as previously emphasized concerning  the right sides of    (\ref{todip4})  and (\ref{top}). However, as stated before, in this paper we are interested in the left sides of those two optical theorems, and hence on the extinctions ${\cal W}_{\hel}^{ext}$ and ${\cal W}^{ext}$, Eqs. (\ref{g}) and (\ref{g1a}), respectively. 

The particle  polarizabilities have a resonance near $\lambda=1.52$  $\mu$m, as shown in Fig. 1, where we have fitted them from their numerical values,  obtained in \cite{silvei}, to functions of $\lambda$; this  enabling us to straightforwardly employ them in Eqs. (\ref{g}) and (\ref{g1a})  . We choose $K=0.6k$, and set $l=1$, $e_0^+ =1$, $e_0^- =i$, i.e.   $\Re\{e_{0}^{-}e_{0}^{+ \, *}\}=0$,  hence  $  \Re \{{\bf E}^{-}\cdot {\bf E}^{+\, *}\}=0$, and so is the third term of (\ref{g1a}) for ${\cal W}^{ext}$;  this  allows to enhancing the value of $g$ even if the incident helicity density  ${\cal H}({\bf r})= (\epsilon/2k)[|{\bf E}^{+}|^2 -|{\bf E}^{-}|^2]$ is very small and the particle were  achiral, as shown below. 

\begin{figure}[htbp]
\centerline{\includegraphics[width=1.1\columnwidth]{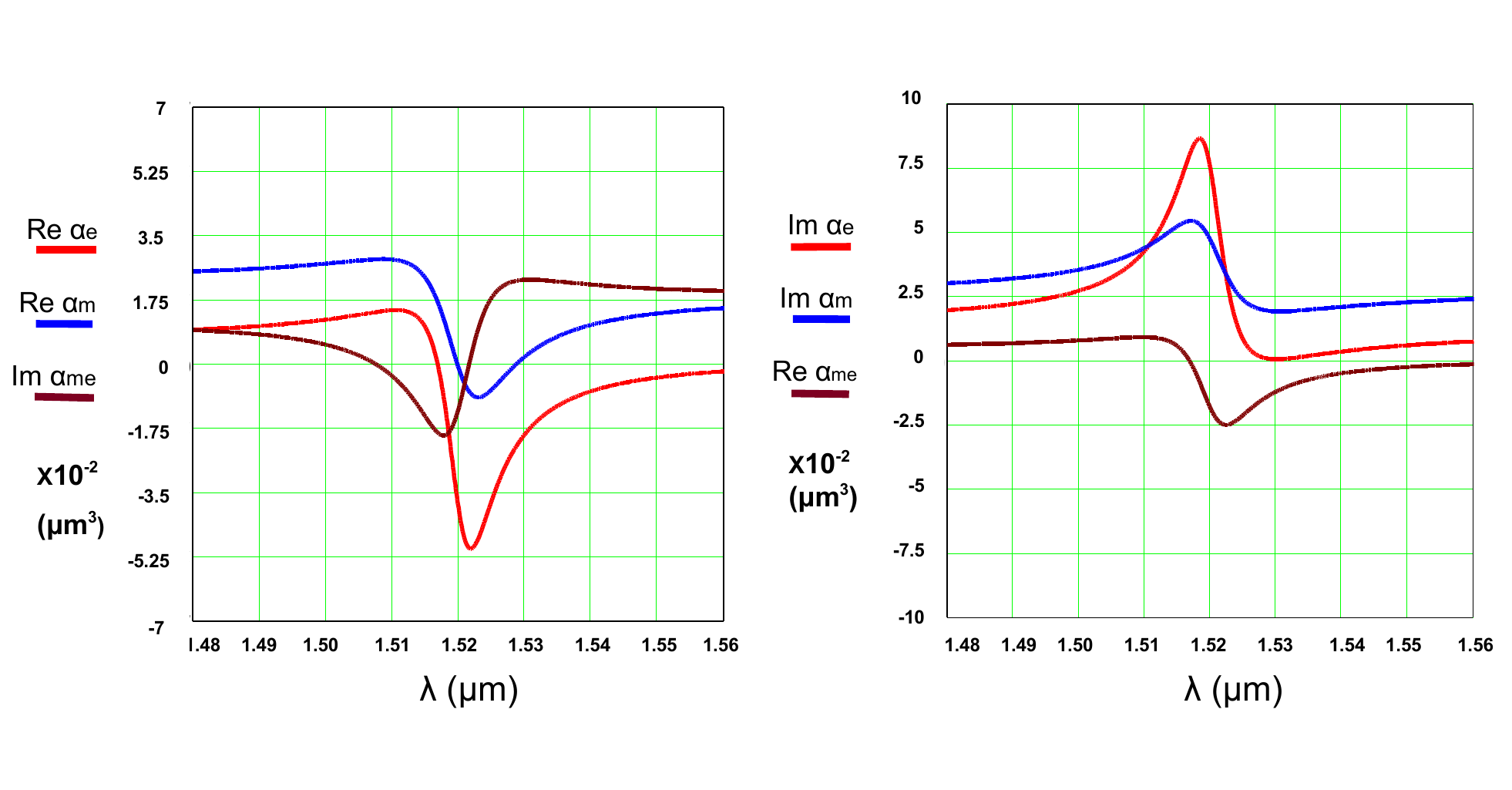}}
\caption{(Color online). Real and imaginary parts of the polarizabilities   $\alpha_e$, $\alpha_m$ and $\alpha_{me}$ of an example of chiral particle near the resonant wavelength   $\lambda=1.52$  $\mu$m. These quantities are functionally fitted from those of  Fig. 4 of \cite{silvei}.  }
\end{figure}

 At the different wavelengths, the radial coordinate $R$ of the particle center in the  beam transversal section is adjusted to two alternative  values of $KR$. One is  $KR\simeq 3.75$, which is near the first out-of axis zero of  $J_{l}^2(KR)=J_1(KR)^2 $, (and thus the contribution of this factor to  Eqs.  (\ref{g}) and (\ref{g1a}) through the first term in the right side of  (\ref{sumdifbess})  is negligible). The other alternative value  is  $KR\simeq 2.25$, which is close to the first zero of $J_{l-1}(KR)=J_0(KR)$, (and hence by virtue of (\ref{sumdif}) the contribution of the $\Im \{ {\bf E}^{-}\cdot {\bf E}^{+\, *}\}$ factor in (\ref{g}) is  negligible). The latter is like the situation of standard CD. 

 These values of  $KR$ also give a hint  on the range $R_0$ of approximate distances   between minima of the beam intensity across its section, versus the size of the particle. $R_0 \simeq 3.75\lambda/2 \pi = 895 nm$ for $\lambda \simeq 1.5 \mu m$, which is  well above the linear size of the particle, which as said above is no larger than $204nm$; and thus allows enough spatial resolution of its position, since this size is well below  the width $R_0$ of the circles of  intensity minima and maxima in the beam section, (see also \cite{volke}).

Hence,   at $KR=2.25$ one has that $|{\bf E}^{+}|^2 +|{\bf E}^{-}|^2=8\pi {\cal W}$  dominates over all other parameters since it is about $2.5 (a.u.)$ while  $\Im \{{\bf E}^{-}\cdot {\bf E}^{+\, *}\}= -0.019$,  $ \Re \{{\bf E}^{-}\cdot {\bf E}^{+\, *}\}=0$, and $|{\bf E}^{+}|^2 -|{\bf E}^{-}|^2= 0.09$. On the other hand, for $KR=3.75$  one sees that  $|{\bf E}^{+}|^2 +|{\bf E}^{-}|^2$ no longer dominates since it is about $0.2 (a.u.)$, while $\Im \{{\bf E}^{-}\cdot {\bf E}^{+\, *}\} =-0.1$,  $|{\bf E}^{+}|^2 -|{\bf E}^{-}|^2=0.008$ and  $\Re \{{\bf E}^{-}\cdot {\bf E}^{+\, *}\}=0$. Therefore these two choices of $kR$ convey a very small incident helicity ${\cal H}$.

Fig. 2 exhibits  the spectra of  the rate of  helicity extinction  ${\cal W}_{\hel}^{ext}$, of energy extinction  ${\cal W}^{ext}$, (both scaled by  $10^{2}/2\pi c)$,  and  helicity extinction  factor $g= {\cal W}_{\hel}^{ext}/{\cal W}^{ext}$, for $KR=3.75$ (upper graph) and  $KR=2.25$ (lower graph) for a chiral  particle with polarizabilities  seen in Fig. 1. We also show  the same scaled quantities, now denoted as  ${\cal W}_{ n\chi \, \hel}^{ext}$, ${\cal W}_{n\chi}^{ext}$, and  $g_{n\chi}$, for an almost achiral particle, ($\alpha_{me} \simeq 0)$, with the same polarizabilities  $\alpha_{e}$ and $\alpha_{m}$ as the former chiral one, but whose cross electric-magnetic polarizability  $\alpha_{me}$ has been somewhat artificially scaled to $1/10$ of the $\alpha_{me}$ values of the chiral particle. (We choose the letter $\chi$ in the subindex  from the Greek $\chi\epsilon\iota\rho$ for "hand").
\begin{figure}[htbp]
\centerline{\includegraphics[width=1.1\columnwidth]{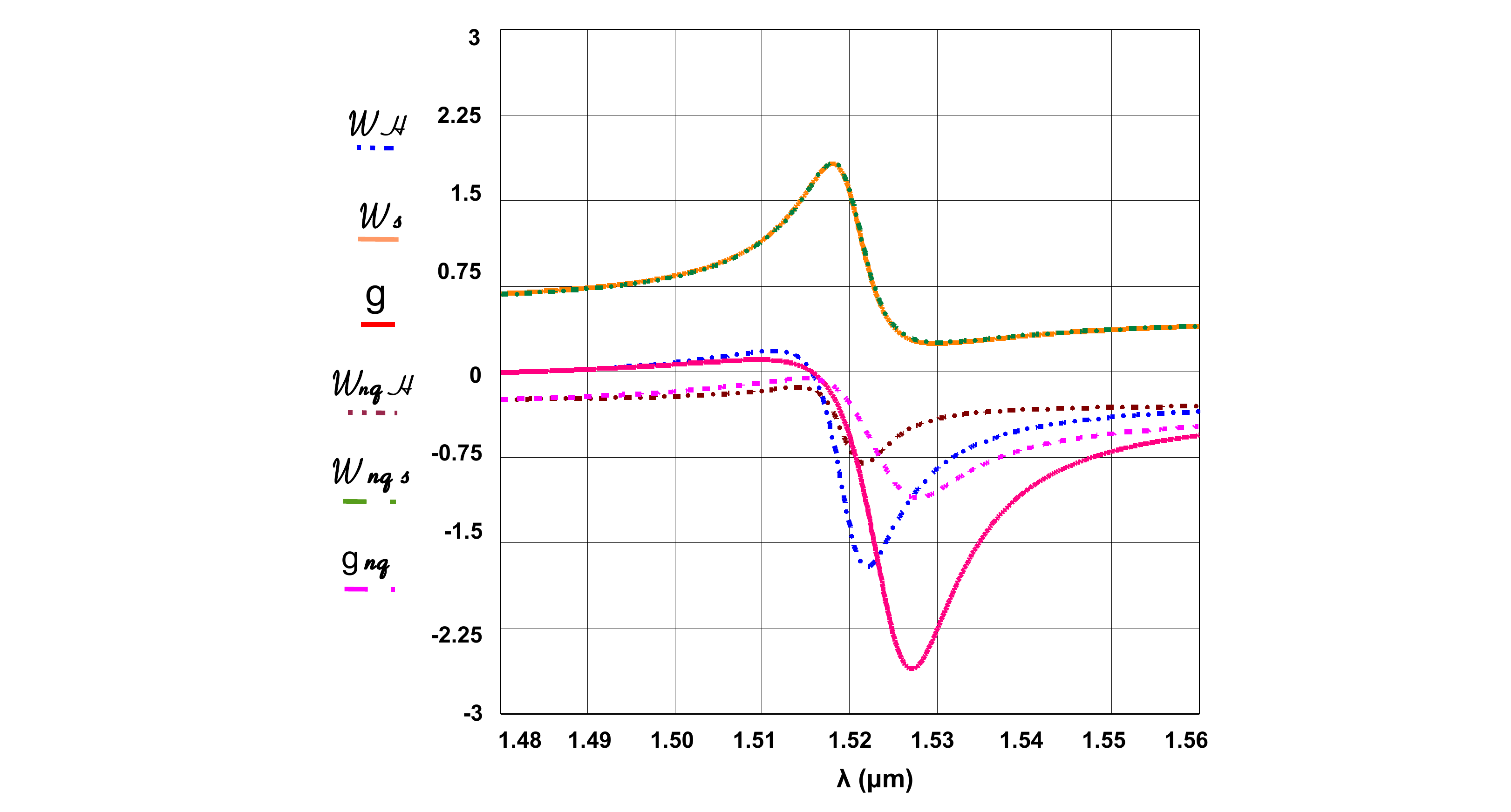}}
\centerline{\includegraphics[width=1.1\columnwidth  ]{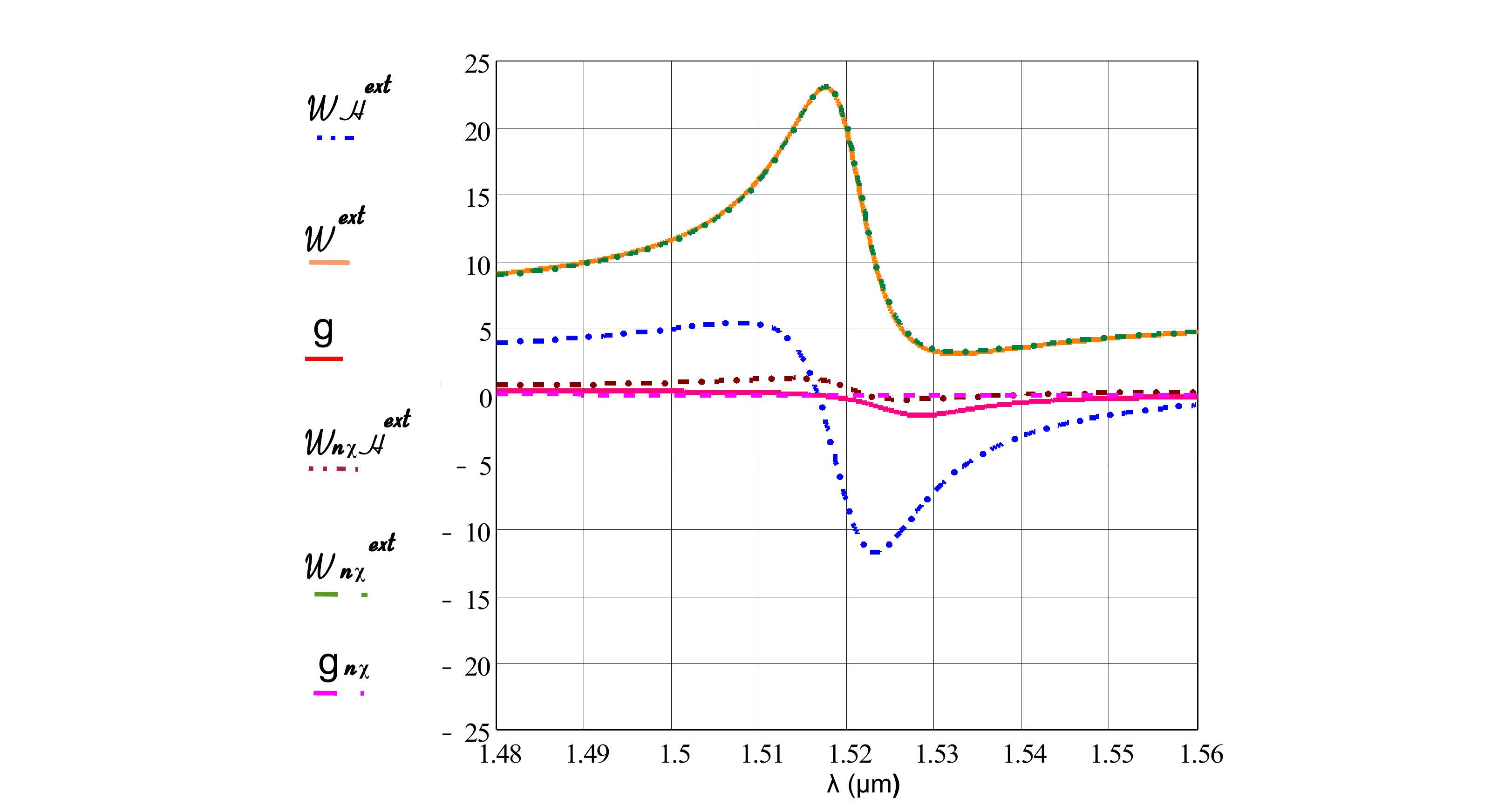}}
\caption{(Color online).  $(10^{2}/2\pi c){\cal W}_{\hel}^{ext}$,  $(10^{2}/2\pi c){\cal W}^{ext}$,  and $g$ for the chiral particle of Figs. 1;  as well as    $(10^{2}/2\pi c){\cal W}_{n\chi \,\hel}^{ext}$,  $(10^{2}/2\pi c){\cal W}_{n\chi}^{ext}$,  and $g_{n\chi}$ when that particle is made achiral, ($\alpha_{me}=0$). Above: $KR=3.75$. Below: $KR=2.25$.}
\end{figure}

As seen from  Eq. (\ref{g}) and Fig. 2 (above), for $KR=3.75$, even  when the particle is  achiral and  the incident helicity density is locally zero, namely at points $R$ fullfiling  $KR=3.75$, {\it we confirm that the interference factor} $ \Im\{ {\bf E}^{-}\cdot {\bf E}^{+\,*}\}$  {\it may be essential to  yield an appreciable  helicity extinction rate} ${\cal W}_{n\chi \,\hel}^{ext}$ {\it and a resonant  helicity  extinction factor}  $|g_{n\chi}| >1$, ($g_{n\chi} =-1.15$ at $\lambda \simeq 1.53 \mu m$ in this illustration). This is one of the main results of this work, and  is in contrast with  standard circular dichroism in which ${\bf E}^{-}\cdot {\bf E}^{+\,*}=0$, and objects with zero, or  a purely imaginary $\alpha_{me}$, with no selective helicity dissipation,  would produce no helicity extinction and therefore  a zero value of $g$ in absence of incident chirality density. 

In this respect we remark that  the appreciable   helicity extinction factors   $g$ and $g_{n\chi}$  observed in Fig. 2 (above), may also be influenced by  shifts, (which depend on the particle morphology), between the resonant peaks of  the  helicity and  energy extinction rates. 

The results of Fig. 2 (above)  should be compared with those  when the factor  $ \Im\{ {\bf E}^{-}\cdot {\bf E}^{+\,*}\}$ is negligible, and so is the third term of Eq. (\ref{g}).    These are plotted in  Fig. 2 (below) for $KR= 2.25$, showing that ${\cal W}_{n\chi \,\hel}^{ext}$ is extremely small compared to ${\cal W}_{n\chi}^{ext}$, and hence  $g_{n\chi}$ is almost zero, ($|g_{n\chi}| \leq 0.05$). This is in contrast with the larger values of   ${\cal W}_{\hel}^{ext}$ and $g$ shown in Fig. 2 (below) for this $KR=2.25$ when the particle  is chiral,  i.e.  $\Re\{\alpha_{me}\}\neq 0$,  and thus in Eq. (\ref{g}) the second term  contributes,   leading to  significantly larger peaks of these quantities,  ($g \simeq  -1.5$ at $\lambda \simeq 1.525 \mu m$), as in standard dichroism.

On the other hand, both Figs. 2 (above and below) show that at a chosen value of $KR$,  ${\cal W}^{ext}$ and ${\cal W}_{n\chi}^{ext}$ coincide with each other; i.e. at a given position of the  particle within the beam,   ${\cal W}^{ext}$  is not affected by the value of $\alpha_{me}$. This is due to the above shown almost negligible  $|{\bf E}^{+}|^2 -|{\bf E}^{-}|^2$, and hence  small ${\cal H}$, for these   chosen $KR$.  Nonetheless when $KR=2.25$,  ${\cal W}^{ext}$, ${\cal W}_{n\chi}^{ext}$ and ${\cal W}_{\hel}^{ext}$ considerably increase  through the factor ${\cal W}$ in (\ref{g}) and (\ref{g1a}). This is expected from the discussion in Section 4 and in this Section 5 above, since  when $KR=2.25$  the factor  $|{\bf E}^{+}|^2 +|{\bf E}^{-}|^2=8\pi {\cal W}$  dominates over all other of Eqs.  (\ref{g}) and  (\ref{g1a}), while  $\Re(\Im) \{{\bf E}^{-}\cdot {\bf E}^{+\, *}\}$ remains  very small, again because  the factor $(K^2 / k_z^2)J_{l-1}(KR) J_{l+1}(KR)$ in (\ref{sumdif}) is much smaller than the first  term proportional to $J_l^2(KR)$ in (\ref{sumdifbess}) when   $KR=2.25$.  Namely, we stress that  in  (\ref{g}) and (\ref{g1a})  $\Re (\Im)\{{\bf E}^{-}\cdot {\bf E}^{+\, *}\}$, Eq.  (\ref{sumdif}), can be either comparable or much smaller  than ${\cal W}$, Eq. (\ref{sumdifbess}), according to the value of $KR$.

 Another consequence of this latter discussion   is that the relative values:    $g/{\cal W}^{ext}$  and $g_{n\chi}/{\cal W}_{n\chi}^{ext}$  are  significantly larger  in Fig. 2 (above) than  in Fig. 2 (below). This once again highlights the relevance of the interference term with factor  $\Re (\Im)\{{\bf E}^{-}\cdot {\bf E}^{+\, *}\}$ in (\ref{g}) and (\ref{g1a}) when the terms proportional to ${\cal W}$  do not dominate.
 
Finally, it should be reminded that, as stated in Section 3,  $\Re\{\alpha_{e}\}$ and $n^2 \Re\{\alpha_{m}\}$ appear substractacted from each other  in the last term of (\ref{g}). Consequently, and although not shown here for brevity, we observe that the amplitude of the peaks of both ${\cal W}_{\hel}^{ext}$  and ${\cal W}_{n\chi \,\hel}^{ext}$ increses  as either  $\Re\{\alpha_{m}\}$  or $\Re\{\alpha_{e}\}$  diminishes. Something analogous occurs with the extinction of energy  (\ref{g1a}) as regards the imaginary parts of the polarizabilities 

\section{Conclusions}
The concept of circular dichroism has been extended  by addressing the  rate of  extinction of helicity  ${\cal W}_{\hel}^{ext}$, whose extinction factor {\it g} has been introduced and generalizes the standard CD dissymmetry factor. The parameter  $g$ monitors the rate of helicity extinction versus that of energy under different values of  the polarizabilities of a generally magnetodielectric particle, either chiral or not, (i.e. for  the cross electric-magnetic one $\alpha_{me}$ ranging from large to almost zero); also considering  the local value of the incident helicity ${\cal H}$. Thus both  ${\cal W}_{\hel}^{ext}$ and {\it g} assess the contibution of the remarkable  interference factor   $\Im \{{\bf E}^{-}\cdot {\bf E}^{+\, *}\}$ to such  helicity extinction in comparison with that of  $\alpha_{me}$, ${\cal H},$ and the resonances of the polarizabilities that we  addressed in this study in order to enhance these effects. Notice in passing that an analogous analysis may be made with the factor $\Re \{{\bf E}^{-}\cdot {\bf E}^{+\, *}\}$ versus  ${\cal W}$, $\alpha_{me},$ and  ${\cal H}$, as regards its contribution to the energy extinction rate    ${\cal W}^{ext}$.

When  the incident fields are optical beams with LCP and RCP transversal components, the factor  $ \{{\bf E}^{-}\cdot {\bf E}^{+\, *}\}$  reduces to that of interference of the  longitudinal components. We have illustrated this with a  Bessel beam. Interestingly,   due to this interference, helicity extinction does not necessarily involve neither particle chirality  nor a non-zero local value of the  incident helicity density; i.e. for  $\alpha_{me}=0$ and  given parameters of the illuminating beam, one may find  positions ${\bf r}_0$ of the particle in the beam where  this local helicity density is ${\cal H}({\bf r}_0)=0$ while the aforementioned interference term gives rise to a non-zero extinction of helicity  ${\cal W}_{\hel}^{ext}$. Also, and importantly,  {\it this interference phenomenon is mediated by  $\Re\{\alpha_{e}\}$ and $\Re\{\alpha_{m}\}$ thus yielding a source of information on these latter quantities, which was not provided by standard CD}.

Finally, although we have  studied these phenomena in general  bi-isotropic dipolar  particles, namely those magnetodielectric  and chiral, the contribution of the $2\Re\{\alpha_{e} -n^{2}\alpha_{m}\}
 \Im\{ {\bf E}^{-} \cdot {\bf E}^{+\,*}\}$  term to an extinction of  incident  helicity, Eqs. (\ref{todip4}) and  (\ref{g}), [as well as the effect of the $2\Im\{\alpha_{e} -n^{2}\alpha_{m}\}
 \Re\{ {\bf E}^{-} \cdot {\bf E}^{+\,*}\}$  term to an extinction of  incident  energy, Eqs. (\ref{top}) and  (\ref{g1a})],  may also be  observed in purely electric ($\alpha_{m}=0$) or magnetic  ($\alpha_{e}=0$) particles. In this context, of special importance will be further research and observation of these effects in high index dielectric particles, that possess  ramarkably unique optically induced electric and magnetic dipole resonances \cite{luki2,nietobol} and that so much interest are generating as  low-loss elements of an increasingly active new area of micro and nano-optics \cite{staude,kivshar2}.

\setcounter{secnumdepth}{0}
\section{Acknowledgments}
Work  supported by MINECO, grants FIS2012-36113-C03-03,  FIS2014-55563-REDC and FIS2015-69295-C3-1-P. The author thanks an anonymous referee for many interesting comments that contributed to improve this report.

\end{document}